\documentclass[twocolumn,amsmath,amssymb]{revtex4}

\def\gsim{ \lower .75ex \hbox{$\sim$} \llap{\raise .27ex \hbox{$>$}} }

\def\lsim{ \lower .75ex \hbox{$\sim$} \llap{\raise .27ex \hbox{$<$}} }

\usepackage{graphicx}
\usepackage{bm}
\input epsf
\def\gsim{ \lower .75ex \hbox{$\sim$} \llap{\raise .27ex \hbox{$>$}} }
\def\lsim{ \lower .75ex \hbox{$\sim$} \llap{\raise .27ex \hbox{$<$}} }

\usepackage{graphicx}
\usepackage{bm}

\newcommand{\be}{\begin{equation}}
\newcommand{\ee}{\end{equation}}
\newcommand{\bea}{\begin{eqnarray}}
\newcommand{\eea}{\end{eqnarray}}
\renewcommand{\d}{\mathrm{d}}

\begin{document}

\title{Dark Energy and the Return of the Phoenix Universe}

\author{Jean-Luc Lehners$^{1}$ and Paul J. Steinhardt$^{1,2}$}
\affiliation{ $^1$Princeton Center for Theoretical Science, Princeton
University, Princeton, NJ 08544 USA \\
$^2$Joseph Henry Laboratories, Princeton
University, Princeton, NJ 08544,
USA
}

\begin{abstract}
In cyclic universe models based on a single scalar field ({\it e.g.}, the
radion determining the distance between branes in M-theory),
virtually the entire universe makes it through the ekpyrotic smoothing
and flattening phase, bounces, and enters a new epoch of expansion and
cooling.
This stable evolution cannot occur, however, if scale-invariant curvature perturbations are produced by the entropic
mechanism  because it
requires two  scalar fields ({\it e.g.}, the radion and the
Calabi-Yau dilaton) evolving along an unstable classical trajectory.
In fact, we show here that an overwhelming fraction of the universe
fails to make it through the ekpyrotic phase; nevertheless, a
sufficient volume survives and cycling continues forever provided the
dark energy phase of the cycle lasts long enough, of order a trillion
years.   Two consequences are a new role for dark energy and a global
structure of the universe radically different from that of eternal inflation.

\end{abstract}


\maketitle

Lemaitre \cite{Lemaitre} invoked the mythical ``phoenix" to refer to
an oscillatory cosmology in which the universe undergoes regular
periods of expansion and contraction with a big crunch/big bang
transition in between.  His model was a closed universe in which the
reversal from expansion to contraction is caused by an overdensity of
matter and radiation.  Although this model is ruled out by WMAP
measurements \cite{Komatsu:2008hk} indicating a flat universe, there
remains the more recent proposal of a cyclic model of the universe
\cite{Steinhardt:2002ih,Steinhardt:2001st} (see \cite{Lehners:2008vx}
for a recent review) that also undergoes regular cycles of evolution
beginning with a big bang and ending in a big crunch.  Unlike
Lemaitre's model, the cyclic universe is kept smooth and spatially
flat by a period of ekpyrotic contraction \cite{Erickson:2003zm,Garfinkle:2008ei} (slow
contraction with equation of state $w \gg 1$) that precedes each big
crunch.  The expansion, contraction and smoothing are governed by a
scalar field $\phi_1$ that evolves along a potential $V(\phi_1)$.  In
a version inspired by M theory \cite{Horava:1995qa}, the cycles
correspond to the regular collision and rebound of two orbifold
planes, and $\phi_1$ is the modulus field whose value determines the
distance between orbifold planes.   As in Lemaitre's oscillatory
model, the notion is that virtually the entire universe makes it
through the big crunch/big bang transition and continues into the next
cycle.

In this paper, we show how the generation of scale-invariant density
perturbations may lead to a variation of the cyclic model that may be
more fittingly called a ``phoenix universe," in the sense that most
of the universe is turned to ``ash" at the end of each cycle and the
universe is reborn from a comparatively tiny
surviving seed.  The phoenix effect is an
unintended byproduct of generating curvature perturbations via the
entropic mechanism \cite{Lehners:2007ac}.  The entropic mechanism was
introduced because an ekpyrotic contraction phase with only a single
scalar field produces scale-invariant time-delay perturbations, but
not scale-invariant curvature perturbations \cite{Creminelli:2004jg}.
Although there are various proposals for circumventing this problem
\cite{Lehners:2007ac,others}, the entropic mechanism is currently the
best understood and can be fully analyzed in four dimensional field
theory without invoking extra dimensions
\cite{Lehners:2007ac,
Tolley:2007nq, Buchbinder:2007tw,Creminelli:2004jg}.
The mechanism
relies on having two  scalar fields (e.g., the radion and the Calabi-Yau dilaton \cite{Lehners:2007ac}) each obtain scale-invariant
fluctuations by evolving down a steep potential $V(\phi_1, \, \phi_2)$
during the ekpyrotic phase, with one linear combination forming a
spectrum of time-delay fluctuations and the second a spectrum of
entropic perturbations.  The latter are converted to curvature
fluctuations if the two-field trajectory bends or bounces as the
universe approaches the big crunch, which occurs generically.

An essential feature of the entropic mechanism is that the
classical ekpyrotic trajectory along $V(\phi_1,\, \phi_2)$
is unstable to transverse
fluctuations \cite{Lehners:2007ac, Tolley:2007nq,
Buchbinder:2007tw,Creminelli:2004jg}.  A consequence
 is that, without adding some mechanism to force the universe to
begin very close to the classical track \cite{Buchbinder:2007tw}, an
overwhelming fraction of the universe fails to make it all the down
the classical trajectory simply due to quantum
fluctuations.  This fraction is transformed into highly
inhomogeneous remnants and black holes that do not cycle or grow in
the post-big bang phase.   However, as shown below, something curious
happens if the dark energy expansion phase preceding the ekpyrotic
contraction phase lasts at least 600 billion years.  Then, a
sufficiently large patch of space makes it all the way down the
classical trajectory and through the big bang such that, fourteen
billion years later, it comprises the overwhelming
majority of space.  This surviving volume,
which grows in absolute size from cycle to cycle, consists of a
smooth, flat, expanding space with nearly scale-invariant curvature
perturbations, in accordance with what is observed today.  As with the
mythical phoenix, a new habitable universe grows from the ashes of the
old.

The phoenix picture is in some sense minimalist.
No features have to
be added to the cyclic model that were not already there, {\it e.g.},
in the original version inspired by M theory and the Ho\v{r}ava-Witten
model
\cite{Horava:1995qa}. One only has had to appreciate the effects of
the elements that were in place.   The heterotic M-theory
embedding of the cyclic model \cite{Lehners:2006pu} already admits a
four-dimensional effective description in terms of gravity coupled to two
canonically
normalized scalar fields.
 The scalars parameterize
geometrical quantities in the higher-dimensional theory, namely
the distance between the boundary branes and the volume of the
internal Calabi-Yau manifold.  Also, dark energy was always
incorporated into the cyclic model to stabilize the oscillatory
behavior.  Here dark energy is given the added role of ensuring the
survival of the phoenix universe. The result is a universe with a
global structure dramatically different from eternal inflation and a
history somewhat different from earlier oscillatory and cyclic
universes.

For concreteness, we will develop the phoenix picture based on
the Ho\v{r}ava-Witten model
\cite{Horava:1995qa} so that there is simultaneously a 4d and higher
dimensional interpretation.  In this picture, the relevant degrees of
freedom aside from the two scalars is the radiation
(with energy density $\rho_\pm$) produced on the
positive and negative tension branes respectively, at the big
crunch/big bang transition; it appears
in the effective theory with a coupling $\beta_\pm$ that
depends on the scalar fields \cite{Steinhardt:2001st}. Thus the
equations of motion are \bea && \!\!\!\!\!\!\!\!\!\! 3H^2 =
\frac{1}{2}(\dot{\phi}_1^2 + \dot{\phi}_2^2) +
V(\phi_i)+\beta_+^4(\phi_i)\rho_+ +\beta_-^4(\phi_i)\rho_-\\ 
&& \!\!\!\!\!\!\!\!\!\! \ddot{\phi}_i +3H\dot{\phi}_i+V_{,\phi_i}=0  \\ &&
\!\!\!\!\!\!\!\!\!\! 3\frac{\ddot{a}}{a} = -(\dot{\phi}_1^2 + \dot{\phi}_2^2) +
V(\phi_i)-\beta_+^4(\phi_i)\rho_+ -\beta_-^4(\phi_i)\rho_-, \eea where we have
assumed a Friedmann-Robertson-Walker background with line
element $\d s^2= -\d t^2 +a^2(t) \d {\bf x}^2;$ $a(t)$ denoting
the scale factor and $H \equiv \dot{a}/a,$ with
$\dot{}\equiv\frac{\partial}{\partial t}.$ These equations form
a closed system subject to the equation of continuity \bea &&
\frac{\partial}{\partial t}(\beta_\pm^4 \rho_\pm) + 4H
\beta_\pm^4 \rho_\pm =0, \eea which implies that \be \beta_+^4
\rho_+ + \beta_-^4 \rho_- = \frac{\rho_0}{a^4}.\ee Hence we can
treat all radiation as one, without the need to know the
precise functional form of $\beta_\pm$.

The effective theory contains a potential that is the sum of
two parts, \be V=V_{ek}(\phi_1,\phi_2)+V_{rep}(\phi_2).\ee The
repulsive potential $V_{rep}$ reflects the fact that there is a
boundary to scalar field space at $\phi_2=0$. This boundary
corresponds to the negative-tension brane being repelled by a
naked singularity in the higher-dimensional bulk spacetime, and
the precise form of the potential depends on the nature of the
matter living on the negative-tension brane ({\it e.g.} scalar matter or a perfect fluid) and on its coupling to the Calabi-Yau volume modulus; see
\cite{Lehners:2006pu,Lehners:2007nb} for details. A simple form
for the repulsive potential is $V_{rep} \propto \phi_2^{-2},$ but we have checked that our
results are also valid for other functional forms of
$V_{rep}$ arising in this way. The ekpyrotic potential is the two-field version of
the cyclic potential described in \cite{Steinhardt:2001st}:
\bea V_{ek}&=&-V_1 e^{-c_1 \phi_1} -V_2 e^{-c_2 \phi_2} +V_0
\eea The essential features are that the potential is steep and
negative, but with a positive plateau at large field values.
Here $V_0,V_i,c_i \gg 1$ are constants, although in general a
small dependence of the $c_i$ on the fields $\phi_i$ is both
natural to expect and necessary to obtain a spectrum with a
small red tilt \cite{Lehners:2007ac}. For the purposes of this
paper we can restrict ourselves to considering constant $c_i.$
During the ekpyrotic phase, the background trajectory is given
by the scaling solution \be a = (-t)^p, \quad \phi_i = {2\over
c_i} \ln (-\sqrt{c_i^2V_i/2} t), \ee where $t=0$ corresponds to
the brane collision and $p \equiv 2/c_1^2 + 2/c_2^2 \ll 1$ is
the equivalent of the slow-roll parameter in inflation. The
trajectory evolves along the ridge in the potential. This is
most easily seen \cite{Koyama:2007mg,Buchbinder:2007tw} by
changing variables to $\sigma$ (pointing along the direction of
the background trajectory) and $s$ (pointing transverse to the
trajectory): \be \sigma \equiv \frac{c_2 \phi_1 +c_1
\phi_2}{\sqrt{c_1^2+c_2^2}} , \qquad s \equiv \frac{c_2 \phi_2
-c_1 \phi_1}{\sqrt{c_1^2+c_2^2}}.\ee Fluctuations in $\sigma$
correspond to adiabatic perturbations, while
fluctuations in $s$ are entropy fluctuations.
During the ekpyrotic phase, the classical background trajectory
must follow the ridge at $s=0,$ and around this trajectory the
potential can be expanded as \bea V_{ek} &=& V_0 -V_0
e^{-\sqrt{2/p}\sigma}(1+\frac{1}{p}s^2+\cdots),
\label{potentialInstability}\eea
where (by a shift in $\phi_1$) we have chosen $V_1$ such that
the prefactor of the second term is $V_0$ also.
\begin{figure}[t]
\begin{center}
\includegraphics[width=0.47\textwidth]{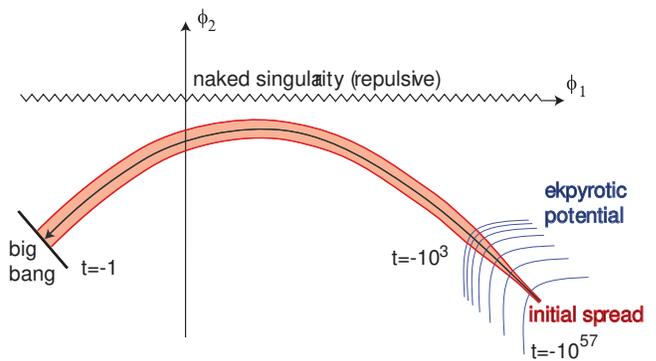}
\caption{\label{FigTrajectory1} {\small This figure shows the
evolution of adjacent classical trajectories from the ekpyrotic phase
through a kinetic phase to the big crunch. The trajectories get
reflected by a boundary in scalar field space at $\phi_2=0.$ All times
are in units of the Planck time. }}
\end{center}
\end{figure}
This expression for the potential clearly shows that the $s$
direction is unstable during ekpyrosis. Hence, unlike for the
single-field cyclic model, the background solution is not an
attractor, and it is not clear at first how
cycling can remain viable. In order to appreciate the severity
of the instability, it is useful to solve the equation of
motion for $s$ in the vicinity of the ridge
\cite{Buchbinder:2007tw}. To leading order in $p$, it is given
by \be \ddot{s}-\frac{2}{t^2}s=0,\ee and subject to the
boundary conditions $s(t_{ek-beg}) \equiv 3s_0/(2t_{ek-beg}),
\dot{s}(t_{ek-beg})=0,$ for some constant $s_0,$ it is solved
by \be s=s_0(\frac{1}{t}+\frac{t^2}{2t_{ek-beg}^3}); \ee in
other words $s$ grows like the Hubble parameter $H.$ Now, the
ekpyrotic phase lasts from $t_{ek-beg}$ to $t_{ek-end} \equiv
e^{-N_{ek}} t_{ek-beg}$ and, hence,
the spread in $s$ grows by a
factor $e^{N_{ek}}.$ $N_{ek}$ denotes the number of e-folds of
ekpyrosis, and in order for the cyclic potential to interpolate
between the dark energy and the grand unified scales (reaching
the grand unified scale is necessary for the perturbations to
have the right amplitude \cite{Lehners:2007ac}), we need
$N_{ek} \approx 120.$ Hence the instability causes $s$ to grow by a huge
factor of about $e^{120}.$ Of course this means that at the
start of the ekpyrotic phase, the field $s$ must be localized
close to the ridge with great precision. As is already
clear from the form of the potential
(\ref{potentialInstability}) and as previously derived in
\cite{Buchbinder:2007tw}, being in the ekpyrotic phase until
$t_{ek-end}$ requires $|s(t_{ek-end})|\lesssim p,$ which
translates into \be |s(t_{ek-beg})|\lesssim p e^{-N_{ek}} 
\label{sLocalization}\ee in Planck units.

\begin{figure}[t]
\begin{center}
\includegraphics[width=0.47\textwidth]{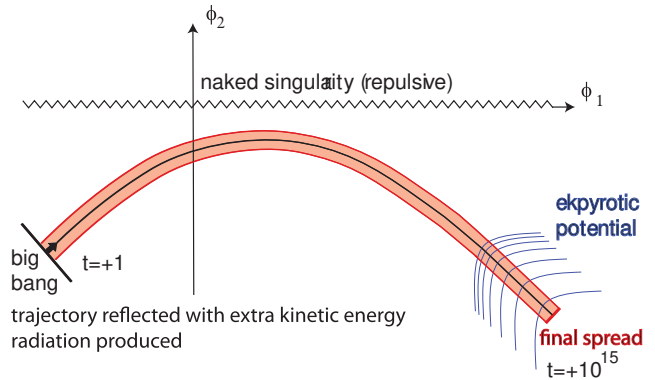}
\caption{\label{FigTrajectory2} {\small After the big bang, adjacent
trajectories proceed through a second kinetic phase and roll back up
the ekpyrotic potential, where the radiation dominated phase starts at
$t\approx 10^{15}$ Planck times. During the ensuing matter and dark
energy phases, the scalar fields remain almost immobile. }}
\end{center}
\end{figure}

After ekpyrosis, the energy density is dominated by the scalar field
kinetic energies. This kinetic phase lasts for about $10^3$
Planck times, during which time the trajectory bends due to the
repulsive potential $V_{rep},$ see Fig. \ref{FigTrajectory1}.
We stop the evolution about one Planck time before the crunch,
and resume the description at about one Planck
time after the big bang, see Fig. \ref{FigTrajectory2}. What we
assume about the crunch/bang transition is that the scalar
field velocities are reversed in direction, and that they are
increased in magnitude by a small parameter $\chi,$ defined as
\cite{Steinhardt:2001st} \be \dot{\phi}_{i,bang} \equiv
-\sqrt{1+\chi}\dot{\phi}_{i,crunch}. \label{definitionChi} \ee
At the brane collision, radiation and matter are produced. In
fact, if there is slightly more radiation produced on the
negative-tension brane than on the positive-tension one, then
the consequence is precisely a slight increase in the scalar
field kinetic energies, as assumed above
\cite{Steinhardt:2001st}. (Some forms of scalar couplings to
the matter fields, $\beta_{\pm}$,
also lead to increased scalar field kinetic
energies \cite{Steinhardt:2001st}.) This extra kinetic energy
helps to overcome the Hubble damping due to the radiation
created at the bang. The constraint that we impose on the
radiation is that it should not come to dominate until the
$\sigma$ field has rolled back up the ekpyrotic potential. As
shown in \cite{Steinhardt:2001st,Khoury:2003rt} this translates
into a constraint on the reheat temperature $T_r$ \be T_r
\leqslant
(-V(t_{ek-end}))^{1/4}|\frac{V_0}{V(t_{ek-end})}|^{\sqrt{3p/16}},\ee
which is easy to satisfy. Finally, we note that the radiation
does not contribute to the repulsive potential $V_{rep}$
\cite{Lehners:2007nb}. However, since there can be extra matter
produced at the bang, which could increase the strength of
$V_{rep},$ we have allowed for a fractional increase in
$V_{rep}$ of (at most) $\chi.$

The extra kinetic energy of the scalars means that the $\sigma$
field quickly rolls back up the unstable ekpyrotic potential to
the (almost-)plateau, and the trajectory deviates little from a
straight line (note that the evolution is still kinetic
dominated at this stage). Numerical simulations show that
during the whole evolution after the ekpyrotic phase, the range
of values of $s$ (which grows by a factor $e^{120}$
during the ekpyrotic phase) increases only by a factor of about
5. Also, the final range encompasses the initial one as long as
$\sqrt{\chi}\lesssim 10^{-2},$ which is a
reasonable assumption.   Numerical simulations
also show that the radiation and increase in $V_{rep}$
discussed above have a very small effect. Hence it is clear now
that there is a fixed point to the entire evolution over the
course of one cycle, and this fixed point is located
exponentially close to the ridge. One might be worried that at
the start of the dark energy phase, when the field finally
turns around on the plateau of the potential, the trajectory
might deviate substantially. This is avoided as long as the
gradient of the potential in the $\sigma$ direction is much
larger than in the $s$ direction, {\it i.e.} we must ensure
that $|V_{,\sigma}/V_{,s}|\approx -\sqrt{p/2}/|s| \gg 1$ or, in
other words, we need to have $|s(t_{ek-beg})|\ll \sqrt{p}.$ On
account of (\ref{sLocalization}), this is clearly satisfied.

So now consider a small range of $s$ around the fixed point:
this range grows by about $e^{N_{ek}}$ over the course of one
cycle, but, crucially, during this time the universe expands by
a factor $e^{2\gamma/3 +N_{rad}+N_{de}},$ where $\gamma \equiv
\ln (|V(t_{ek-end})|^{1/4}/T_r)$ parameterizes the total growth
during the kinetic phases and $N_{rad} \equiv \ln (T_r/T_0)$ is
the number of e-folds of expansion between the start of
radiation domination and today. Hence if \be N_{de} > N_{ek}
-\frac{2}{3}\gamma -N_{rad} \approx 60, \ee then the gradient
of values of $s$ is sufficiently diluted that a small initial
flat patch will grow into a larger flat patch after each cycle. In
other words, an initial patch of space, with a spread in field values
$\Delta s$ around the fixed point value, will be amplified over the
course of one cycle to a region containing a larger patch of space
with the same spread $\Delta s.$
Hence, starting from an arbitrarily small initial flat patch
with the right values of $s$ one can cycle to a large, flat and
habitable universe.

In principle, cycling can be eternal to the past and future; in this
case, the patch of surviving space in a phoenix universe
would be infinite all along and
would increase in volume by an exponential factor each cycle.  An
alternative possibility is that the universe had a definite beginning: for example if it was born in an initial
quantum creation event, such as in the Hartle-Hawking no boundary
proposal \cite{Hartle:1983ai}, or, in the colliding branes picture,
through the sudden quantum creation from nothing of a positive- and
negative-tension orbifold plane pair with random, but smooth, initial
conditions.  All that is required is that there be a finite
probability for a patch to have the right values of $s$ to grow
from cycle to cycle.  Even if the branes and/or patch begin small, one ends up with
an arbitrarily large, smooth, flat, habitable universe full of matter and radiation
as cycles proceed.

\begin{figure}[t]
\begin{center}
\includegraphics[width=0.47\textwidth]{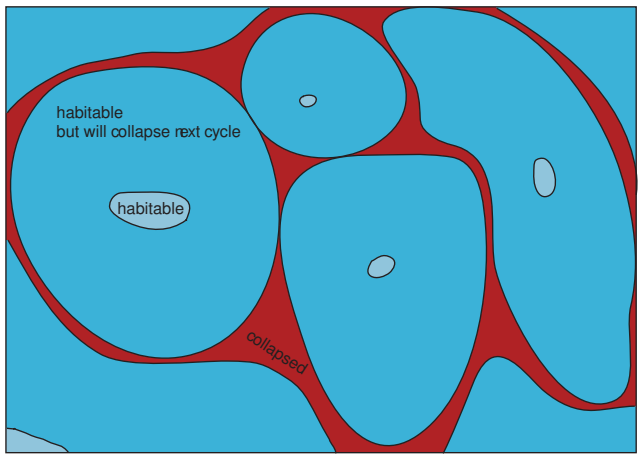}
\caption{\label{FigGlobal} {\small The global structure of the two-
field cyclic universe: large smooth and flat regions are interspersed with
small regions that have collapsed and have stopped cycling. The tiny regions of lighter shade will turn into the entire habitable regions during the next cycle. }}
\end{center}
\end{figure}

The global structure of the phoenix universe is quite different from
that of eternal inflation.  First of all, there is no amplification of rare
quantum fluctuations, as there is for inflation.  The phoenix universe, like
general cyclic models, has the
property that the smoothing phase (dark energy followed by ekpyrotic
contraction) has a smaller expansion rate, {\it i.e.} a smaller Hubble
parameter $H$, than the  matter and radiation dominated phases. Hence,
if rare quantum fluctuations keep some region in the smoothing phase while typical regions
proceed to the matter and radiation dominated phases, the rare regions lose in volume
to the typical ones.  The situation is reversed with inflation: rare regions kept  behind
in the smoothing (de Sitter) phase are exponentially enhanced and
soon dominate the volume.   This
principle underlies eternal inflation.  The result is
that most of the volume in an inflationary
universe is empty, uninhabitable and exponentially expanding, punctuated by rare bubbles where inflation
has ended, some (but not all) of which are habitable.  By contrast, most of the volume
in a cyclic universe is habitable.

The phoenix universe adds a new wrinkle because of the instability associated
with the entropic mechanism.  Namely, only a tiny fraction of the
habitable volume in any given cycle survives to the next cycle.  Most of the volume
fails to make it through the ekpyrotic smoothing phase
and collapses through a mixmaster crunch \cite{Garfinkle:2008ei} and/or
black hole formation (see also \cite{Erickson:2006wc}).  Nevertheless, because the small
surviving fraction grows in volume by an exponential factor after the bounce (and the failed regions
collapse), most of space during the cycle remains habitable.  In this
sense, habitability is an attractor in the phoenix universe.
Of course, to some extent the present discussion pertains more to
metaphysics than to physics - the important question is whether
the model discussed here leads to any distinct observational
consequences.

One observational signature of great promise has been discussed
recently: due to the very steep ekpyrotic potential, ekpyrotic
and cyclic models distinguish themselves by a substantial and
soon measurable amount of non-Gaussianity
\cite{Koyama:2007if,Buchbinder:2007at,Lehners:2007wc}. A second
potentially observable effect is suggested by the above
analysis in the case that the cyclic universe has a
definite beginning: since the dark energy phase has only just begun,
there should still be a significant spread of values of $s$
across the currently observable universe. Hence one would
expect the dark energy not to be equal everywhere in space, but
to vary by a fractional amount of \be \frac{\Delta V}{V}
\approx \frac{V_{,ss}(\Delta s)^2}{V} \approx \frac{(\Delta
s)^2}{p}.\ee Since the spread in values of $s$ satisfies
$\Delta s \lessapprox p$ and since $10^{-4} \lesssim p \lesssim
10^{-2}$ for models that give observationally acceptable levels
of spectral tilt and non-Gaussianity, we would expect a spatial
variation in dark energy of \be \frac{\Delta V}{V}\lessapprox
10^{-4}.\ee  One would expect to be closer to the upper
bound if we are still in one of the early cycles,
 and less and
less variation as we get to older cycles. The largest variation
occurs for a flat region that was smaller than the horizon
size in previous cycles and which exceeds the present
horizon size for the first time
during the current cycle. However, there is no particular
reason to believe that we must be in one of the very early cycles.
Nevertheless, conceptually the spatial
variation of dark energy can be regarded as a cosmic clock
in scenarios where the cyclic universe has a beginning.

Finally, we wish to emphasize the importance of dark energy to the phoenix universe.  In previous studies of the cyclic universe based on a single field where there was no instability during the ekpyrotic phase,
the role of dark energy was confined to ensuring
that the cyclic solution was a stable attractor solution to the equations of motion
in the case that the radion overshoots
after a bounce (or the branes separate too much after
a collision). Dark energy is not needed to smooth or
flatten the universe; this is easily accomplished during the ekpyrotic contraction phase alone.  To meet the attractor
requirement, only two or three e-folds of dark energy domination are needed. For the phoenix universe, though, dark energy plays a new role: it expands the region of space that meets the conditions on $s$ required to complete the unstable ekpyrotic trajectory and rebound as a smooth, flat patch with nearly scale-invariant fluctuations in accord with what is observed.  With too few e-folds of dark energy, this patch would shrink from one cycle to the next and would not survive.  With at least 60 e-folds of accelerated expansion at the current rate (or  600 billion years of dark energy phase), the patch grows from cycle to cycle, and the phoenix universe is forever reborn.  Dark energy is, thereby, given a new role and must satisfy a new constraint.  The new constraint is rather mild quantitatively, but, qualitatively, by controlling the survival or termination of the phoenix universe, it may act as a selection criterion that may help to explain why the dark energy density is so small and yet non-zero today \cite{future}.

We would like to thank Jim Peebles and Neil Turok for stimulating
discussions and comments on the manuscript.
This work was supported in part by US Department of Energy grant DE-FG02-91ER40671.

\end{document}